\documentclass[a4paper,11pt]{article}
\pdfoutput=1
\usepackage{jheppub}
\usepackage{braket}
\usepackage{bm}
\usepackage{subcaption}
\usepackage{xspace}

\newcommand{\Libra}{\texttt{Libra}\xspace}
\newcommand{\LiteRed}{\texttt{LiteRed}\xspace}
\newcommand{\e}{\epsilon}
\newcommand{\eell}{\ensuremath{e^+e^-\to \ell^+\ell^-}\xspace}
\newcommand{\eemumu}{\ensuremath{e^+e^-\to \mu^+\mu^-}\xspace}
\newcommand{\cA}{\mathcal{A}}
\newcommand{\cT}{\mathcal{T}}
\newcommand{\cAg}{\mathcal{A}_{1\gamma}}
\newcommand{\cAgg}{\mathcal{A}_{2\gamma}}
\newcommand{\cAggg}{\mathcal{A}_{3\gamma}}
\newcommand{\cH}{\mathcal{H}}

\title{Electron-positron annihilation into heavy leptons at two loops.}
\author{Roman E. Gerasimov,}
\author{Petr A. Krachkov,}
\author{and Roman N. Lee}

\affiliation{Budker Institute of Nuclear Physics, Novosibirsk 630090, Russia}

\emailAdd{r.e.gerasimov@inp.nsk.su}
\emailAdd{p.a.krachkov@inp.nsk.su}
\emailAdd{r.n.lee@inp.nsk.su}

\abstract{We calculate the NNLO QED corrections to the $C$-even part of differential cross section of $e^+e^- \to \mu^+\mu^-$ process. We neglect power corrections in the electron mass and obtain the result in terms of Goncharov's polylogarithms.}

\begin{document}

\maketitle
\flushbottom

\section{Introduction}

Muon-antimuon pair production in electron-positron annihilation is one of the most fundamental QED processes. Its amplitude in the Born approximation is a perfect starting point for first acquaintance with perturbative calculations which is witnessed by many textbooks on quantum field theory. What is more important, this process plays a crucial role in any experiment with electron-positron colliders. Besides being used for determining luminosity, it also gives a large background for more rare events including possible manifestations of New Physics. The Born approximation is not sufficient for these purposes and one needs to take into account higher orders of perturbation theory in the fine structure constant $\alpha\approx 1/137.036$. The NLO corrections to this process have been calculated long ago in Refs. \cite{Berends1983,Jadach1984}. For the precision of the planned collider experiments the NNLO precision is required, see, e.g., recent review \cite{Aliberti:2024fpq} and references therein.

We stress that the exact account of the produced lepton mass $m_\ell$ ($\ell=\mu,\tau$) is important not only for the production of $\tau$-leptons, but also for that of muons. In particular, this account is required for the correct measurement of the cross section of $e^+e^-\to \pi^+\pi^-$. The measurement of this cross section is of great importance, especially at relatively low energies, as it contributes significantly to hadron vacuum polarization correction to $g_\mu-2$. As the muon mass is of the same order of magnitude as that of pion, the exact account for this mass is necessary both for precise luminosity measurements and for the background separation.

In contrast, the electron mass $m_e$ can always be considered as a small parameter. Even for muon production on the threshold the power corrections with respect to this parameter are suppressed as $(m_e/m_\mu)^2\sim 2.3\times 10^{-5}$ and can be safely neglected. In particular, in Refs. \cite{Berends1983,Jadach1984} these power corrections have been omitted. However, due to the collinear divergences which appear in the amplitude with massless electron, one can not simply put $m_e$ to zero. Note that for QCD calculations one usually does put the light quark mass to zero. But this is entirely due to the fact, that the collinear divergences disappear when the hard cross sections are convoluted with parton distribution functions of \emph{colorless} (\emph{zero color charge}) hadrons. Meanwhile, in QED processes with the \emph{charged} particles involved the collinear divergences must be tamed by the account of those particles' masses, even if they are small, to give mass logarithms.

The important prerequisite of NNLO calculation is the availability of the corresponding two-loop master integrals. In Refs.\cite{mastrolia2017master,Becchetti:2019tjy} the two-loop master integrals for the $e\mu\to e\mu$ process were calculated at zero electron mass. Note that this process is a cross-channel of \eell. However the analytical continuation to the annihilation channel is not trivial due to the lack of Euclidean kinematic region\footnote{Confer Ref. \cite{tausk1999non}, where the expressions for on-shell massless nonplanar double box in different channels can not be obtained by analytical continuation due to the same reason. In that paper the result for different regions was obtained by considering the double box with one off-shell leg, so that the parameters $s$, $t$, $u$ were independent and the Euclidean region $s,t,u<0$ existed.} as explained in Ref \cite{LM2019}, where a subset of the two-loop master integrals for the \eemumu process was calculated. Nevertheless, in Ref. \cite{Bonciani:2021okt} such an analytical continuation was done numerically and the contribution to differential cross section from the spin-averaged interference of two-loop amplitudes with the Born amplitude for massless electron was obtained. As expected, the result contained soft and collinear divergences. Meanwhile, as we explain above, the observable cross sections for light charged particles do not contain the collinear divergences but necessarily contain the mass logarithms.

The goal of the present paper is to provide the analytical results for the two-loop QED amplitudes of the \eell process which contribute to $C$-even part of the differential cross section. We show that for this contribution the electron mass should be taken into account only in the one- and two-loop form factors and polarization operator, which are known, with some reservations, exactly in this parameter for a long time. In what follows, we will talk about the process \eemumu for definiteness, but the same consideration with some obvious modifications is also valid for the production of $\tau$-leptons.

\section{General consideration}

We consider the process
\begin{equation}
	e^-(p_1)+e^+(p_2) \longrightarrow \mu^-(q_1)+\mu^+(q_2)\,.
\end{equation}
and introduce conventional invariants
\begin{equation}
	s=(p_1+p_2)^2=(q_1+q_2)^2\,,\quad
	t=(p_1-q_1)^2=(p_2-q_2)^2\,,\quad
	u=(p_1-q_2)^2=(p_2-q_1)^2\,.
\end{equation}
The momenta and invariants satisfy usual constraints
\begin{equation}
	p_1+p_2=q_1+q_2\,,\quad
	p_1^2=p_2^2=m^2\,,\quad
	q_1^2=q_2^2=M^2\,,\quad
	s+t+u =2m^2+2M^2\,,
\end{equation}
where $m=m_e$ and $M=m_\mu$ are the electron and muon masses, respectively.
All through the paper we will also use the notation $\beta = \sqrt{1-\tfrac{4M^2}{s}}$ which is the muon velocity in c.m. frame.

We will use the following prescription for the dimensionally regularized loop integration:
\begin{equation}
	\frac{d^4l}{(2\pi)^4} \to
	\frac{\underline{d}^dl}{(2\pi)^d}  =
	\left(\frac{e^{\gamma_E}}{4\pi}\right)^{\e}  \frac{d^dl}{(2\pi)^d} =
	\frac{1}{(4\pi)^2}\frac{e^{\e\gamma_E} d^dl}{\pi^{d/2}}\,,
	\label{eq:loop_measure}
\end{equation}
where $\gamma_E=0.577\ldots$ is the Euler constant, $d=4-2\e$. Consequently, in the formulae below we will use notations
\begin{equation}
	\underline{d}^dl=\left(\tfrac{e^{\gamma_E}}{4\pi}\right)^\e d^dl\,,\qquad
	\underline{d}^{d-1}l=\left(\tfrac{e^{\gamma_E}}{4\pi}\right)^\e d^{d-1}l\,.
\end{equation}
We use dimensional regularization to regularize both the ultraviolet and infrared divergences.

\begin{figure}
	\includegraphics[width=\textwidth]{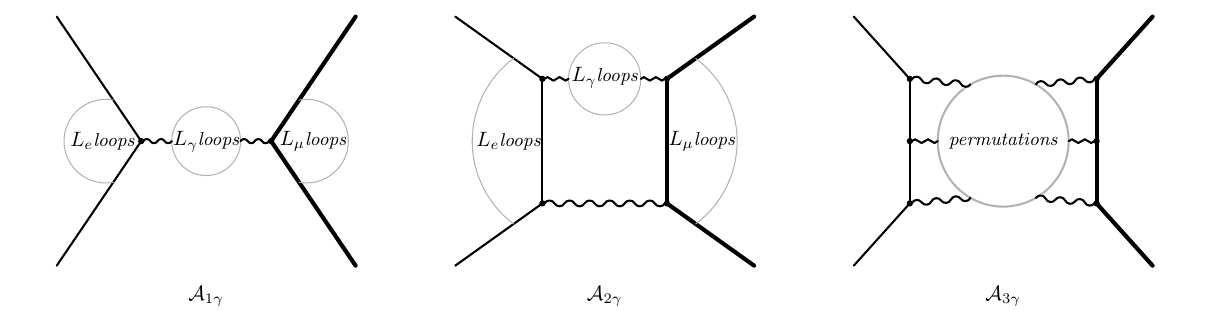}
	\caption{$1\gamma$-, $2\gamma$- and $3\gamma$-reducible diagrams that contribute to \eemumu process up to two loops. First set contains one- and two-loop corrections to the electron and muon form factors and to the photon self-energy, $L_e+L_\mu+L_\gamma \leqslant2$. In the second diagram $L_e+L_\gamma+L_\mu\leqslant 1$.
	}
	\label{fig:NNLOdiagrams}
\end{figure}

The diagrams which contribute to the \eemumu amplitude, in addition to the number of loops $L$, can be conveniently graded by the minimal number $n$ of the photon lines which should be cut to split the diagram into two parts, one containing initial electron and positron, another --- final muon and antimuon. We will call such a diagram $n\gamma$-reducible. Obviously, we have diagrams with $L\geqslant n-1 \geqslant 0$. Thus, the full amplitude reads
\begin{equation}
	\cA=\sum_{L=0}^{\infty} a^{L+1}\sum_{n=1}^{L+1}  \cA_{n\gamma}^{(L)}\,,
\end{equation}
where $a=\frac{\alpha}{4\pi}$ ($\alpha\approx1/137.036$ is the fine structure constant) and we denote by $a^LA_{n\gamma}^{(L)}$ the $L$-loop $n\gamma$-reducible part of the amplitude.

Up to two loops we have $1\gamma$-, $2\gamma$-, and $3\gamma$-reducible diagrams, see Fig. \ref{fig:NNLOdiagrams}. The diagrams with two photons describe the production of muons in $C$-even state, while those with one or three photons correspond to the muon production in $C$-odd state. In the total cross section the interference between $C$-odd and $C$-even diagrams vanishes. While it is not so in the differential cross section, the interference between $C$-odd and $C$-even diagrams is an odd function of $c=\cos \theta$, where $\theta$ is a scattering angle. Therefore, in the experimental setup it is usually possible to separate the $C$-even part of the differential cross section from the $C$-odd one. The former comes from the interference of $C$-odd diagrams with $C$-odd ones or from that of $C$-even diagrams with $C$-even ones.

Up to two loops the cross section reads
\begin{align}
	d\sigma&=\left|\cA\right|^2d\Phi=d\sigma_{C\text{-even}}+d\sigma_{C\text{-odd}},\\
	d\sigma_{C\text{-even}}&=a^2\Big[|\cA_{1\gamma}^{(0)}|^2
	+2\Re \cA_{1\gamma}^{(1)} \cA_{1\gamma}^{(0)*}a
	+\Big(|\cA_{1\gamma}^{(1)}|^2+|\cA_{2\gamma}^{(1)}|^2\nonumber\\&
	\qquad\qquad\qquad+2\Re \cA_{1\gamma}^{(2)} \cA_{1\gamma}^{(0)*}
	+2\Re \cA_{3\gamma}^{(2)} \cA_{1\gamma}^{(0)*}\Big)a^2
	\Big]d\Phi,\label{eq:dsigma_even}\\
	d\sigma_{C\text{-odd}}&=2a^3\Re \left[
	\cA_{2\gamma}^{(1)} \cA_{1\gamma}^{(0)*}
	+\left(\cA_{2\gamma}^{(1)} \cA_{1\gamma}^{(1)*}
	+\cA_{2\gamma}^{(2)} \cA_{1\gamma}^{(0)*}\right)a
	\right]d\Phi\,.
\end{align}
Our normalization corresponds to the expression
$$
\cA_{1\gamma}^{(0)} = \frac1s \overline{U}(q_1) \gamma^{\mu} V(q_2) \,\bar v(p_2) \gamma_{\mu} u(p_1)
$$
for the Born amplitude, so that the phase space $d\Phi$ reads (we omit $\sqrt{1-4m_e^2/s}$ in the denominator)
\begin{equation}
	d\Phi =\frac8{s^2}\, \delta(p_1+p_2-q_1-q_2)\, \frac{\underline{d}^{d-1}{q}_1\,\underline{d}^{d-1}{q}_2}{(2\pi)^{d-6}}\,.
\end{equation}

In the present paper we will consider only the contributions to $C$-even part of the differential cross section, which is, therefore, an even function of $c=\cos \theta$. The contributions to the $C$-odd part of the cross section (charge asymmetry) will be considered in a separate paper.

Note that some  diagrams with the insertion of hadronic vacuum polarization also contribute at relative orders $\alpha$ and $\alpha^2$. Although these contributions can not be calculated from the first principle, they can, in principle, be expressed as a weighted integral of imaginary part of hadronic vacuum polarization function $\Pi_{\text{hadr}}(s)$. For the $C$-even part of the differential cross section up to NNLO these are the $1\gamma$-reducible diagrams only. We present the corresponding formulae when discussing the  vacuum polarization and form factors contribution.

\subsection{Collinear divergences and account of electron mass}
\begin{figure}
	\includegraphics{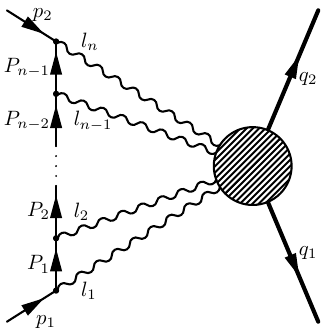}
	\centering
	\caption{Gauge-invariant sum of diagrams corresponding to Eq. \eqref{eq:S}.}
	\label{fig:cur}
\end{figure}
We are now in position to analyze which diagrams develop collinear divergence at zero electron mass. Let us consider the gauge-invariant sum of diagrams (see Fig. \ref{fig:cur}) with massless electron which can be represented in the form
\begin{equation}\label{eq:S}
S=\int d^dl_1\ldots d^dl_{n}\delta\left(p_1+p_2-\sum_{k=1}^{n} l_k\right)\frac{j^{\mu_n\ldots \mu_1}J_{\mu_n\ldots \mu_1}}{{P_{n-1}^2}\ldots {P_{1}^2}l_n^2\ldots l_1^2}\,.
\end{equation}
Here
\begin{equation}
	j^{\mu_1\ldots \mu_n}=\bar{v}\gamma^{\mu_n}{\widehat{P}_{n-1}}\gamma^{\mu_{n-1}}\ldots{\widehat{P}_{1}}\gamma^{\mu_{1}}u,\qquad P_k=p_1-\sum_{r=1}^{k}l_r,
\end{equation}
and $J_{\mu_n\ldots \mu_1}$ is a ``conserved current'' satisfying
\begin{equation}
	 l_k^{\mu_k} J_{\mu_n\ldots \mu_1}= 0,\qquad (k=1,\ldots, n)
	 \label{eq:transv}
\end{equation}
The collinear divergences may come from the integration region where $l_1,\ldots, l_k\raisebox{-1.5mm}{$\overset{\propto}{\sim}$} \,p_1$ and/or $l_r,\ldots, l_n\raisebox{-1.5mm}{$\overset{\propto}{\sim}$} p_2$ with $1\leqslant k < r\leqslant n$, where $\raisebox{-1.5mm}{$\overset{\propto}{\sim}$}$ denotes approximate collinearity.
However, it is easy to check that in this region the numerator in Eq. \eqref{eq:S} is suppressed due to the Dirac equations $\hat{p}_1 u=\bar v \hat{p}_2=0$ and transversality condition \eqref{eq:transv}. Therefore, in such diagrams the collinear divergences do not appear.

These diagrams can be conveniently described as the ones which do not contain a photon line which connects two points on the electron line. We stress, that the above argumentation concerns the whole gauge-invariant set of the diagrams, which guarantees the condition \eqref{eq:transv} to be fulfilled. The individual diagrams described as above may still exhibit collinear divergences, but they cancel in the gauge-invariant sum.

Note that the diagrams containing a photon line which connects two points on the electron line can not be represented in the form \eqref{eq:S} with $J_{\mu_n\ldots \mu_1}$ satisfying Eq. \eqref{eq:transv}. Such are the diagrams in Fig. \ref{fig:NNLOdiagrams} from sets $\cA_{1}$ and $\cA_{2}$ with $L_e>0$. Therefore, only for those diagrams we have to keep the electron mass nonzero.
A good news from the above consideration is that we already have all ingredients for the calculation of $d\sigma_{C\text{-even}}$ at NNLO.\footnote{Note, that the master integrals of the set $\cAgg$ in Fig. \ref{fig:NNLOdiagrams} with $L_e=1$ with the account of the electron mass were recently calculated in Ref. \cite{Lee:2024jvd} thus paving a way to the calculation of $d\sigma_{C\text{-odd}}$ at NNLO.} Indeed, the diagrams of the set $\cAg$ are  known exactly in the masses of both particles. The two-loop diagrams of the set $\cAgg$ do not enter $d\sigma_{C\text{-even}}$. Finally, the diagrams of the set $\cAggg$ are expressed in terms of zero electron mass master integrals calculated in Ref. \cite{LM2019}.

\subsection{Soft divergences}
In the previous subsection we have discussed the collinear divergences of the amplitudes. But there also soft divergences. In a seminal paper \cite{Yennie1961} it was shown that these divergences factorize in amplitude as
\begin{equation}
	\cA = e^{a\mathcal{V}} \cH\,,\label{eq:AviaH}
\end{equation}
where $\cH$ is the so called hard massive amplitude, which is finite at $d=4$, and the factor $e^{a\mathcal{V}}$ absorbs all soft singularities. The function $\mathcal{V}$ is defined as\footnote{Note that, in contrast to QCD, in QED the perturbative expansion in $a$ of $V$ contains only one-loop contribution.}
\begin{equation}
\mathcal{V}=-\sum_{i<j}Q_{i}Q_{j}V\left(p_{i},p_{j}\right),
\end{equation}
where the sum runs over all pairs of incoming and outgoing particles, all momenta $p_{i}$ are considered to be incoming, and $Q_{i}$ is the charge, in units of $|e|$, of $i$-th particle multiplied by $\pm1$ for incoming/outgoing particle, respectively. The function $V\left(p_{i},p_{j}\right)$ is defined as\footnote{Note, that for convenience, we have introduced the factor $\left(\tfrac{e^\gamma}{4\pi}\right)^{\e}$ compared to the definition of Ref. \cite{Fadin:2023phc}.}
\begin{equation}
V\left(p_{i},p_{j}\right)=
-8\pi^2
\int\frac{\underline{d}^{d}k}{i\left(2\pi\right)^{d}}\frac{1}{k^{2}+i0}\left(\frac{2p_{i}-k}{k^{2}-2kp_{i}+i0}+\frac{2p_{j}+k}{k^{2}+2kp_{j}+i0}\right)^{2}\,.\label{eq:SVfunction}
\end{equation}
The explicit form of this function up to $\e^0$ can be found in the Appendix \ref{sec:SV}.

For our present setup we have
\begin{align}\label{eq:Vif}
	\mathcal{V}&=\mathcal{V}_{II}+\mathcal{V}_{FF}+\mathcal{V}_{IF}
	=V(p_1,p_2)+V(-q_1,-q_2)+2[V(p_1,-q_1)-V(p_1,-q_2)].
\end{align}
The advantage of using the hard amplitudes $\cH$ is that we can put $\e=0$ in them. One might argue that without knowing the higher-order $\e$-expansion terms in $\cH$ we will not be able to reconstruct the amplitude $\cA$ via Eq. \eqref{eq:AviaH}. This is true, but we still will be able to use the hard amplitudes to obtain the observable differential cross section inclusive with respect to soft photons. Namely, it is well known \cite{Bloch:1937pw,Yennie1961} that the radiation of soft photons factorizes in such a cross section and we have
\begin{equation}
	d\sigma_{\text{incl}}(\omega_0) = e^{a\mathcal{W}(\omega_0)}d\sigma,
\end{equation}
where $d\sigma=\left|\cA\right|^2d\Phi$ is the ``elastic'' cross section, $\omega_0$ is the maximal energy of each soft photon.
The function $\mathcal{W}(\omega_0)$ reads
\begin{equation}\label{eq:W}
	\mathcal{W}(\omega_0)  =  -\sum_{i<j}Q_iQ_jW(p_i,p_j|\omega_0)\,,
\end{equation}
where
\begin{equation}
	W(p_i,p_j|\omega_0)=-16\pi^2\intop_{\omega<\omega_0}\frac{\underline{d}^{d-1}k}{(2\pi)^{d-1}2\omega}\left(\frac{p_i}{k\cdot p_i}-\frac{p_j}{k\cdot p_j}\right)^2 \label{eq:Wint}
\end{equation}
The explicit form of slightly differently defined soft real function was obtained in Ref. \cite{Lee:2020zpo} up to $\e^0$. For reader convenience we present the corresponding formulas for our present definition \eqref{eq:Wint} in Appendix \ref{sec:SV}.

For our present setup, similar to Eq. \eqref{eq:Vif} we have
\begin{align}\label{eq:Wif}
	\mathcal{W}&=\mathcal{W}_{II}+\mathcal{W}_{FF}+\mathcal{W}_{IF}
	=W(p_1,p_2)+W(-q_1,-q_2)+2[W(p_1,-q_1)-W(p_1,-q_2)].
\end{align}
The important point is that the divergence-free inclusive cross section is an observable, in contrast to the ``elastic'' cross section.
Written in terms of hard amplitude, it reads
\begin{equation}
	d\sigma_{\text{incl}}(\omega_0) = e^{a\mathcal{W}(\omega_0)+2a\Re \mathcal{V}}\left|\cH\right|^2d\Phi.
	\label{eq:csviaH}
\end{equation}
Now, although each function $\mathcal{W}(\omega_0)$ and $2\Re \mathcal{V}$ contains divergences, their sum is finite at $\e=0$. Thus we conclude that all three quantities, $\cH$, $\mathcal{W}(\omega_0)$ and $\mathcal{V}$, are sufficient to be known up to $\e^0$ in order to construct the observable inclusive cross section.

The decomposition \eqref{eq:Vif} allows us to define the finite hard contributions $a^l \cH_{n\gamma}^{(l)}$ such that
\begin{align}
	a \cA_{1\gamma}^{(0)}+a^2 \cA_{1\gamma}^{(1)}+a^3 \cA_{1\gamma}^{(2)}+O(a^4)
	&= e^{a(\mathcal{V}_{II}+\mathcal{V}_{FF})}\left[a \cH_{1\gamma}^{(0)}+a^2 \cH_{1\gamma}^{(1)}+a^3 \cH_{1\gamma}^{(2)} + O(a^4)\right]\\
	a \cA_{1\gamma}^{(0)}+a^2 \cA_{2\gamma}^{(1)}+a^3 \cA_{3\gamma}^{(2)}+O(a^4)
	&= e^{a\mathcal{V}_{IF}}\left[a \cH_{1\gamma}^{(0)}+a^2 \cH_{2\gamma}^{(1)}+a^3 \cH_{3\gamma}^{(2)} + O(a^4)\right]
\end{align}
These two equations define amplitudes $\cA_{1\gamma}^{(0)}$, $\cA_{1\gamma}^{(1)}$, $\cA_{2\gamma}^{(1)}$, $\cA_{1\gamma}^{(2)}$, $\cA_{3\gamma}^{(2)}$ which enter the $C$-even cross section \eqref{eq:dsigma_even} in terms of the corresponding hard amplitudes $\cH_{1\gamma}^{(0)}$, $\cH_{1\gamma}^{(1)}$, $\cH_{2\gamma}^{(1)}$, $\cH_{1\gamma}^{(2)}$, $\cH_{3\gamma}^{(2)}$ and vice versa. Let us write the explicit relations
\begin{align}
	\cH_{1\gamma}^{(0)} &=\cA_{1\gamma}^{(0)}, \nonumber\\
	\cH_{1\gamma}^{(1)} &= \cA_{1\gamma}^{(1)}-\big(\mathcal{V}_{II}+\mathcal{V}_{FF}\big)\cA_{1\gamma}^{(0)},\nonumber\\
	\cH_{2\gamma}^{(1)} &= \cA_{2\gamma}^{(1)}-\mathcal{V}_{IF}\cA_{1\gamma}^{(0)},\nonumber\\
	\cH_{1\gamma}^{(2)} &= \cA_{1\gamma}^{(2)} - \big(\mathcal{V}_{II}+\mathcal{V}_{FF}\big)\cA_{1\gamma}^{(1)}
	+\tfrac12\big(\mathcal{V}_{II}+\mathcal{V}_{FF}\big)^2\cA_{1\gamma}^{(0)},\nonumber\\
	\cH_{3\gamma}^{(2)} &= \cA_{3\gamma}^{(2)}-\mathcal{V}_{IF}\cA_{2\gamma}^{(1)}+\tfrac12\big(\mathcal{V}_{IF}\big)^2\cA_{1\gamma}^{(0)}\label{eq:HviaA}
\end{align}
In each of thus defined quantities the soft singularities cancel.
In the $C$-odd cross section there enters one more amplitude, $\cA_{2\gamma}^{(2)}$. Let us present for completeness the definition of the corresponding hard amplitude $\cH_{2\gamma}^{(2)}$:
\begin{align}
	\cH_{2\gamma}^{(2)} &= \cA_{2\gamma}^{(2)}-\mathcal{V}_{IF}\cA_{1\gamma}^{(1)}-\left(\mathcal{V}_{II}+\mathcal{V}_{FF}\right)\cA_{2\gamma}^{(1)}+\mathcal{V}_{IF}\left(\mathcal{V}_{II}+\mathcal{V}_{FF}\right)\cA_{1\gamma}^{(0)}\,.
\end{align}
It is obvious that the expressions of $\cA$ in terms of $\cH$ are given by the same formulae with the replacement $\cH\leftrightarrow\cA$, $\mathcal{V}_{\bullet}\to -\mathcal{V}_{\bullet}$.

Moreover, it is easy to see that the cancellation of soft divergences can be further specialized. Namely, if we introduce notation $\cA_{n\gamma,L_e,L_\gamma,L_\mu}^{(L)}$ to denote the diagrams with fixed parameters $L_e,L_\gamma,L_\mu$, see Fig. \ref{fig:NNLOdiagrams}, we can define, e.g.
\begin{align}
	\cH_{1\gamma,1,0,1}^{(2)} &= \cA_{1\gamma,1,0,1}^{(2)} - \mathcal{V}_{II}\cA_{1\gamma,0,0,1}^{(1)}-\mathcal{V}_{FF}\cA_{1\gamma,1,0,0}^{(1)}
	+\mathcal{V}_{II}\mathcal{V}_{FF}\cA_{1\gamma,0,0,0}^{(0)},\\
	\cH_{2\gamma,1,0,0}^{(2)} &= \cA_{2\gamma,1,0,0}^{(2)}-\mathcal{V}_{IF}\cA_{1\gamma,1,0,0}^{(1)}-\mathcal{V}_{II}\cA_{2\gamma,0,0,0}^{(1)}+\mathcal{V}_{IF}\mathcal{V}_{II}\cA_{1\gamma,0,0,0}^{(0)}\,.
\end{align}
which are also finite at $\e=0$.

\section{Tensor decomposition}

Since our goal is to obtain the NNLO cross section \emph{including} the polarization effects, we have to determine the linearly independent (spin-)tensor structures which may appear in the amplitude. Then the coefficients in front of those tensor structures, the invariant amplitudes, will determine the differential cross section for any polarization state of the particles. We will follow the most radical approach by constructing the full set of possible tensor structures in $d$ dimensions. The same approach was used in Ref. \cite{Fadin:2023phc}, which lead the authors to 7 tensor structures of which one was evanescent in $d=4$. For our present four-fermion amplitude we observe an infinite hierarchy of tensor structures which is constructed of antisymmetric products of $\gamma$-matrices for each of the two fermion chains contracted with each other over Lorentz indices. More precisely, we have the following basis
\begin{align}
	\cT_{n,j,i}&= \overline{U}  T_{j,\mu_1...\mu_n} V \,\bar v t_i^{\mu_1,...,\mu_n} u,\nonumber\\
	t_0^{\mu_1...\mu_n}&=\gamma^{[\mu_1}\ldots\gamma^{\mu_n]}\,,\quad
	t_1^{\mu_1...\mu_n}=(q_1-q_2)_\mu \gamma^{[\mu}\gamma^{\mu_1}\ldots\gamma^{\mu_n]}\,,\nonumber\\
	T_0^{\mu_1...\mu_n}&=\gamma^{[\mu_1}\ldots\gamma^{\mu_n]}\,,\quad
	T_1^{\mu_1...\mu_n}=(p_1-p_2)_\mu\gamma^{[\mu}\gamma^{\mu_1}\ldots\gamma^{\mu_n]}\,,\label{eq:ts}
\end{align}
where
\begin{equation}
	\gamma^{[\mu_1}...\gamma^{\mu_n]}={\frac1{n!}}\sum_{\sigma\in S_n} (-1)^{|\sigma|}\gamma^{\mu_{\sigma_1}}\ldots \gamma^{\mu_{\sigma_n}}
\end{equation}
denotes the anti-symmetrized product of matrices and $u = u(p_1),\,v=v(p_2),\, U=U(q_1),\,V=V(q_2)$ are the Dirac spinors of electron, positron, muon, and antimuon, respectively.

In what follows we will enumerate these tensor structures using one index as follows
\begin{equation}
	\cT_{n,j,i}=\cT_{4n+2j+i}
\end{equation}
Given the variation range of $i$ and $j$ indices, see Eq. \eqref{eq:ts}, it is obvious that this notation is unambiguous.
Despite the fact that the basis of tensor structures is infinite, at each loop level only a finite subset of structures $\cT_k$ may appear as the maximal number of $\gamma$-matrices in each fermion chain is restricted. In particular, for two loops there may be up to $5$ gamma-matrices in each fermion chain. Consequently, there are only 21 tensor structures $\{\cT_0,\ldots, \cT_{20}\}=\{\cT_{0,0,0},\ldots, \cT_{5,0,0}\}$ which may appear. Up to two loops the amplitude can be expanded as
\begin{equation}
	\cA=\sum_{k=0}^{20} A_k \cT_k,
\end{equation}
where the scalar functions $A_k$ are the invariant amplitudes. In order to find them, we calculate the matrix
\begin{equation}
	M_{k,k'}=\langle \cT_k\cT_{k'}^*\rangle,
\end{equation}
where $\langle \ldots\rangle$ denotes the averaging over the spin variables:
\begin{multline}
	\langle \overline{U} \Gamma_1 V\,\bar v \Gamma_2 u(\overline{U} \Gamma_3 V\,v \Gamma_4 u)^*\rangle\\
	=\mathrm{Tr}[(\hat{q}_1+M)\Gamma_1(\hat{q}_2-M)\gamma^0\Gamma_3^{\dagger}\gamma^0]
	\times\mathrm{Tr}[(\hat{p}_2-m)\Gamma_2(\hat{p}_1+m)\gamma^0\Gamma_4^\dagger\gamma^0]/(\mathrm{Tr}\,1)^{2}\,.
\end{multline}

Then we have
\begin{equation}
	A_k = \sum_{k'}\left\langle\cA \cT_{k'}^{*}\right\rangle \left(M^{-1}\right)_{k'k}
\end{equation}
The invariant amplitudes $A_k$ determine the differential cross section for any polarization states of the particles. In particular, the unpolarized cross section reads
\begin{equation}
	d\sigma = 4A_k M_{kk'}A_{k'}^* d\Phi
\end{equation}

\subsection{Invariant amplitudes at $d=4$.}

Of course, at $d=4$ we have yet fewer possible tensor structures.
In particular, the tensor structures with more than 4 antisymmetrized gamma-matrices vanish identically. This leaves us with $17$ tensor structures of which, due to 4-dimensional algebra of gamma-matrices, only $8$ are linearly independent. Those eight structures can be chosen as
\begin{equation}
	\{\cT_0,\ldots,\cT_7\}\,,\label{eq:basis4}
\end{equation}
while the remaining elements can be represented as
\begin{align}
  \cT_{8}&=-\frac{1}{2(M^4-t u)}\left((s-4M^2)(t-u)\cT_0
  +2M(s-4M^2)\cT_2
  +(t-u)\cT_7\right)\,,\nonumber\\
  \cT_{9}&=\frac{2M}{t-u}\cT_{11}=-\frac{2M}{3}\cT_{12}=-\frac{4M}{s}\left(\cT_3
  +(t-u)\cT_4\right)\,,\nonumber\\
  \cT_{10}&=\frac{M}{(M^4-t u)}\left((t-u)^2\cT_0+2M(t-u)\cT_2+s \cT_7\right)\,,\nonumber\\
   \cT_{15}&=\frac{u-t}{4} \cT_{16}=\frac{3}{2}\frac{t-u}{M^4-tu}\left((t-u)^2\cT_0+ 2M(t-u)\cT_2+ s \cT_7\right) \,,\nonumber\\
   \cT_{13}&=\cT_{14}=\cT_{17}=\cT_{18}=\cT_{19}=\cT_{20}=0\,.
\end{align}
In the above relations we have neglected the electron mass.
The number of elements in \eqref{eq:basis4} can be easily understood as the number of independent helicity amplitudes $\mathfrak{H}_{\lambda_{e-}\lambda_{e+}\lambda_{\mu-}\lambda_{\mu+}}$ when the $P$-parity is conserved. Namely, we have
\begin{gather}
	\mathfrak{H}_{++++}\stackrel{P}{=}\mathfrak{H}_{----},\quad
	\mathfrak{H}_{+++-}\stackrel{P}{=}\mathfrak{H}_{---+},\quad
	\mathfrak{H}_{++-+}\stackrel{P}{=}\mathfrak{H}_{--+-},\quad
	\mathfrak{H}_{++--}\stackrel{P}{=}\mathfrak{H}_{--++},\\
	\mathfrak{H}_{+-++}\stackrel{P}{=}\mathfrak{H}_{-+--},\quad
	\mathfrak{H}_{+-+-}\stackrel{P}{=}\mathfrak{H}_{-+-+},\quad
	\mathfrak{H}_{+--+}\stackrel{P}{=}\mathfrak{H}_{-++-},\quad
	\mathfrak{H}_{+---}\stackrel{P}{=}\mathfrak{H}_{-+++}
\end{gather}
Then, if we take into account also the $C$-parity conservation, we have additional identities
\begin{gather}
	\mathfrak{H}_{+++-}\stackrel{C}{=}	\mathfrak{H}_{++-+},\quad
	\mathfrak{H}_{+-++}\stackrel{C}{=}\mathfrak{H}_{+---}\,,
\end{gather}
which leaves us with 6 independent amplitudes
\begin{gather}
	\mathfrak{H}_{++++}, \quad \mathfrak{H}_{+++-},\quad \mathfrak{H}_{++--},\\
	\mathfrak{H}_{+-++}, \quad \mathfrak{H}_{+-+-},\quad \mathfrak{H}_{+--+}\,.
\end{gather}
Finally, for massless electron we have helicity conservation requirement which forbids the amplitudes on the first row of the above equation thus leaving us with three amplitudes.

Luckily, our choice of tensor structures $\cT_k$ has secured that we observe exactly three nonzero coefficients $A_1,\, A_3,\, A_4$ in our results in front of
\begin{align}
	\cT_1 &= \overline{U}V \,\bar v (\hat q_1-\hat q_2) u,\\
	\cT_3 &= \overline{U}(\hat p_1-\hat p_2)V \,\bar v (\hat q_1-\hat q_2) u,\\
	\cT_4 &= \overline{U}\gamma_{\mu}V \,\bar v\gamma^{\mu}u.
\end{align}

We finish this Section by noting that all above considerations are equally valid for hard amplitudes $\cH$. In particular,
\begin{equation}
	\cH= H_1 \cT_1 +H_3 \cT_3 + H_4 \cT_4\,.
\end{equation}

\section{$1\gamma$-reducible diagrams}
We can represent the contribution of $\cAg$ diagrams in Fig. \ref{fig:NNLOdiagrams} in the form
\begin{align}\label{eq:A1g}
\cAg=\frac{a}{s \left[1-\Pi_{tot}(s)\right]}\overline{U}\left(\Gamma^\mu\right)_{(\mu)} V \bar{v}(\overline{\Gamma}_{\mu})_{(e)} u\,,
\end{align}
Here $\Pi_{tot}(s)$ is one-photon irreducible polarization operator and the vertices $\left(\Gamma^\mu\right)_{(e,\mu)}$ are expressed via form factors $F_{k,(e,\mu)}$ as
\begin{align}\label{formactor}
\left(\Gamma^\mu\right)_{(\mu)}&=\gamma^\mu F_{1,(\mu)}(s)-\frac{\sigma^{\mu\nu}P_\nu}{2M} F_{2,(\mu)}(s),\\
\left(\overline{\Gamma}_\mu\right)_{(e)}&=\gamma_\mu F_{1,(e)}(s)+\frac{\sigma_{\mu\nu}P^\nu}{2m} F_{2,(e)}(s),
\end{align}
where  $P=p_1+p_2=q_1+q_2$, so that $P^2=s$, and $\sigma^{\mu\nu}=\gamma^{[\mu}\gamma^{\nu]}=\tfrac12[\gamma^{\mu}\gamma^{\nu}-\gamma^{\nu}\gamma^{\mu}]$.

The  electron Pauli form factor $F_{2,(e)}(s)\sim O(m^2/s)$ can be neglected within our precision.

Using the identity
\begin{equation}
	\overline{U}\gamma^{[\mu}\gamma^{\nu]}P_\nu V =
	\overline{U}\left[-2M \gamma^\mu+q_1^\mu-q_2^\mu\right] V\,,
\end{equation}
we obtain
\begin{align}\label{eq:A1gA}
	\cAg&=A_{1,1\gamma} \cT_1 + A_{4,1\gamma} \cT_4,\\
	A_{1,1\gamma}&=-\frac{a F_{1,(e)}F_{2,(\mu)}}{2M s \left[1-\Pi_{tot}(s)\right]}
	\\
	A_{4,1\gamma}&=\frac{aF_{1,(e)}(F_{1,(\mu)}+F_{2,(\mu)})}{s \left[1-\Pi_{tot}(s)\right]}
\end{align}

Note that the tensor structure $\cT_3$ does not appear in $1\gamma$-reducible contributions. It is easy to understand from the fact that such diagrams have definite parity (negative) with respect to charge conjugation of electron-positron, or muon-antimuon, wave functions separately.

\subsection{Polarization operator}
The polarization operator can be written as a sum of lepton and hadron parts
\begin{equation}
	\Pi_{tot}(q^2)=\sum_{\ell}\Pi_{\ell}(q^2)+\Pi^{(\text{had})}(q^2)\,.
\end{equation}
We define the coefficients $\Pi_{\ell}^{(l)}$ of the perturbative expansion of $\Pi_{\ell}$ as
\begin{equation}\label{eq:Pi_pert}
	\Pi_{\ell}(q^2)=\sum_l a^l\Pi_{\ell}^{(l)}(q^2)\,.
\end{equation}
Up to two loops the leptonic polarization operator is universal, i.e.
\begin{equation}
	\Pi_{\ell}(q^2)=\Pi(q^2/m_{\ell}^2)\,,
\end{equation}
where the function $\Pi(x)$ does not depend on the flavor.
Two-loop polarization operator $\Pi(x)$ has been known since the work of Kallen and Sabry \cite{KallenSabry}. The simplest form was recently presented in Ref. \cite{PhysRevD.109.096020}. The asymptotics for $x\gg 1$ can be easily obtained from the known result.

Concerning the (renormalized) hadronic polarization operator $\Pi^{(\text{had})}(s)$, it is at present not possible to obtain it from the first principles. Its imaginary part $\Im \Pi^{(\text{had})}(s)$ is related to the observable inclusive hadronic cross section $e^+e^-\to\text{hadrons}$ and real part can be obtained via the dispersion relation
\begin{equation}
	\Pi^{(\text{had})}(q^2)=\frac{q^2}{\pi}\intop_{s_0}^{\infty} \frac{ds\, \Im \Pi^{(\text{had})}(s)}{s(s-q^2)}\,.
	\label{eq:dispersion}
\end{equation}
In the leading order in $\alpha$ the relation between $\Im \Pi^{(\text{had})}(s)$ and the standard quantity
\begin{equation}
	R(s)=\frac{\sigma_{e^+e^-\to\text{hadrons}}}{\sigma_{e^+e^-\to\mu^+\mu^-,\text{Born}, m_\mu=0}}
\end{equation}
reads
\begin{equation}
	 \Im \Pi^{(\text{had})}(s) = - \frac{\alpha}{3} R(s)\,.\label{eq:Pi2R}
\end{equation}
However, for our present purpose we need $\Pi^{(\text{had})}$ with the account of the first correction in $\alpha$. With this precision its relation with what is called $\sigma_{e^+e^-\to\text{hadrons}}$ in the experiments needs to be scrutinized. We do not consider this question in the present paper.

\subsection{Form factors}
Similar to the polarization operator, the form factors can also be represented as
\begin{equation}
	F_{k,(\ell),\text{tot}}(q^2)=
	{F}_{k,(\ell)}(q^2)+F_{k,(\ell)}^{(\text{had})}(q^2)\,.\label{eq:Ftot}
\end{equation}
where the term $F_{k,(\ell)}^{(\text{had})}$ stands for the contribution of the diagram with hadronic vacuum polarization insertion.

We define the coefficients $F_{k,(\ell)}^{(l)}$ of the perturbative expansion of the form factors via
\begin{equation}
F_{k,(\ell)}=\sum_L
a^L
F_{k,(\ell)}^{(L)}.\label{eq:F_pert}
\end{equation}
Note that we use dimensional regularization to regularize both the ultraviolet and the infrared divergences. While the former disappears after renormalization, the latter still remains. Therefore, when defining the perturbative expansion we should pay attention to the factors which tend to unity when $\epsilon$ goes to zero. In particular, our definition of dimensionally regularized loop integration measure \eqref{eq:loop_measure} differs from that used in Ref. \cite{Mastrolia2003,Bonciani2004}. One should remember this fact when making the comparison of the results.
\begin{figure}
	\centering
	\includegraphics[width=1\linewidth]{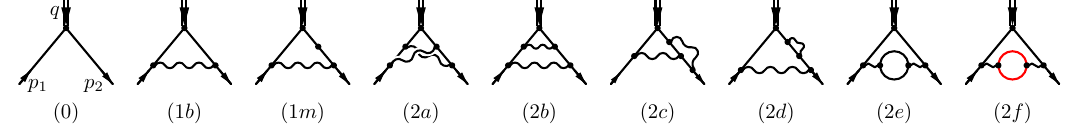}
	\caption{Diagrams for unrenormalized form factors up to two loops.
		The external legs are labeled as on the first diagram. The last diagram depicts the contribution with another lepton loop. The diagram $(1m)$ is required to construct the renormalized form factors. The dot on the fermion line on this diagram stands for the fermion-fermion vertex $-im$.}
	\label{fig:ff}
\end{figure}
We have independently obtained the expansion of the form factors up to two loops, including the contribution of the two-loop diagram with another species lepton loop. The corresponding diagrams are depicted in Fig. \ref{fig:ff}.

We have found an agreement with the results of Refs. \cite{Mastrolia2003,Bonciani2004}. Note, however, that the published expressions in those papers contain typos both in the results for the master integrals and in the results for the renormalized form factors and we thank the authors for sending us their results in an electronic form which we actually used to make a comparison.
Concerning the contribution to the form factors of the lepton loop of another species, we have obtained the expressions valid for all kinematic regions. In Ref. \cite{Ahmed2024} this contribution has been considered in one kinematic region. We have compared our results for this region with those of Ref. \cite{Ahmed2024} and found an agreement apart from the opposite sign. Details of our calculation of this contribution  are presented in the Appendix \ref{sec:vp_vertex}.

Let us define the ``hard'' form factors $F_{\cH,k,(\ell)}$ as
\begin{gather}
	F_{\cH,1,(e)}=e^{-a\mathcal{V}_{II}}F_{1,(e)},\quad
	F_{\cH,2,(e)}=e^{-a\mathcal{V}_{II}}F_{2,(e)},\\
	F_{\cH,1,(\mu)}=e^{-a\mathcal{V}_{FF}}F_{1,(\mu)},\quad
	F_{\cH,2,(\mu)}=e^{-a\mathcal{V}_{FF}}F_{2,(\mu)}
\end{gather}

Then from Eqs. \eqref{eq:A1gA} and \eqref{eq:AviaH} we have the following expressions for the contribution of $1\gamma$-reducible diagrams into hard invariant amplitudes $H_1$ and $H_4$:
\begin{align}
		H_{1,1\gamma}&=-\frac{a F_{\cH,1,(e)}F_{\cH,2,(\mu)}}{2M s \left[1-\Pi_{tot}(s)\right]}
	\\
	H_{4,1\gamma}&=\frac{aF_{\cH,1,(e)}(F_{\cH,1,(\mu)}+F_{\cH,2,(\mu)})}{s \left[1-\Pi_{tot}(s)\right]}
\end{align}
\begin{figure}
	\centering
	\includegraphics[width=0.25\linewidth]{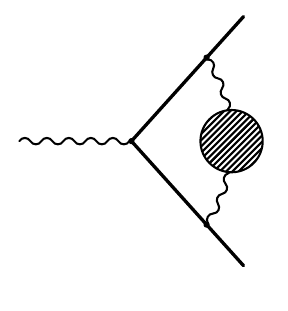}
	\caption{Contribution of hadronic vacuum polarization to form factors.}
	\label{fig:hvpvertex}
\end{figure}
The term $F_{k,(\ell)}^{(\text{had})}$ in Eq. \eqref{eq:Ftot}, which stands for the contribution of the diagram in Fig. \ref{fig:hvpvertex}, can not be calculated from the first principles. However, using Eq. \eqref{eq:dispersion} we can represent them as
\begin{equation}
	F_{k,(\ell)}^{(\text{had})}(s)=-\frac{4a}{\pi}  \int_{s_0}^{\infty } \, \frac{ds_1 }{s_1} K_{k}\left(\tfrac{s}{m_\ell^2},\tfrac{s_1}{m_\ell^2}\right)\Im\Pi^{(\text{had})}(s_1)\label{eq:hadr_vert}
\end{equation}
where the integral kernels $K_{1,2}$ are expressed in terms of the dilogarithm function. Their explicit form is presented in Appendix \ref{sec:hadr_vert}.

Note that within our precision, in these contributions we can use Eq. \eqref{eq:Pi2R} to obtain the numerical estimate for $F_{k,(\ell)}^{(\text{had})}$ from the experimental data for $\sigma_{e^+e^-\to\text{hadrons}}$.

\section{$2\gamma$- and $3\gamma$-reducible diagrams}

As we already stated above, the contributions $\cA_{2\gamma}^{(1)}$ and $\cA_{3\gamma}^{(2)}$ which enter the $C$-even part of the cross section do not develop collinear divergences and, therefore, can be calculated at zero electron mass. Fortunately, the master integrals for this setup have been calculated in Ref. \cite{LM2019}.
We use these results with some modifications.

In Ref. \cite{LM2019} the variables $\tilde{x}$ and $\tilde{z}$ nave been introduced.\footnote{Therein these variables were called $x$ and $z$, respectively, but we add tilde here to avoid confusion with the notations of the present paper.} They are related to $s$ and $t$ via
\begin{equation}
	\frac{s}{M^2}=\left(\tilde{x}+\frac{1}{\tilde{x}}\right)^2\,\quad \frac{t}{M^2} = 1-\frac{\left(\tilde{x}+\frac{1}{\tilde{x}}\right)^2}{\tilde{z}^2+1}\,.
\end{equation}
The master integrals obtained in  Ref. \cite{LM2019} are expressed in terms of Goncharov's polylogarithms  $G(\boldsymbol{a}|\tilde{z}-\tilde{x}^{-1})$ with $a_k\in \{-\tilde{x}^{-1},\pm \tilde{x}^{\pm1}-\tilde{x}^{-1},\pm i-\tilde{x}^{-1},\pm (\tilde{x}^2+1+\tilde{x}^{-2})^{\pm1/2}-\tilde{x}^{-1},\pm i(\tilde{x}^2+3+\tilde{x}^{-2})^{1/2}-\tilde{x}^{-1}\}$.

In these variables the physical region corresponding to the annihilation channel is defined by inequalities
\begin{equation}
	\tilde{x}>1\,,\quad \tilde{x}^{-1}<\tilde{z}<\tilde{x}\,.
\end{equation}
The forward-backward symmetry with respect to the replacement $t\leftrightarrow u$ acts on these variables as
$\tilde{z}\leftrightarrow \tilde{z}^{-1}$. Both the arguments and the parameters of the Goncharov's polylogarithms in the results of Ref. \cite{LM2019} are modified nontrivially by this substitution which we consider as somewhat inconvenient.

Therefore we have undertaken some efforts to pass to more symmetric variables. Namely, we introduce

\begin{equation}
	\beta = \frac{\tilde{x}^2-1}{\tilde{x}^2+1},\quad c=\cos\theta=\frac{\left(\tilde{x}^2+1\right) \left(\tilde{z}^2-1\right)}{\left(\tilde{x}^2-1\right) \left(\tilde{z}^2+1\right)}\,,
\end{equation}
where $\beta$ is the velocity of the produced leptons in c.m.s. and $\theta$ is the scattering angle.

While these variables are sufficient for the one-loop and planar two-loop master integrals, for the nonplanar master integrals we had to introduce another pair of variables,
\begin{equation}
	\xi=\frac{\tilde{x}-1}{\tilde{x}+1} = \tfrac{\sqrt{1+\beta}-\sqrt{1-\beta}}{\sqrt{1+\beta}+\sqrt{1-\beta}}\,,\quad
	\chi = \frac{(\tilde{x}+1) (\tilde{z}-1)}{(\tilde{x}-1)(\tilde{z}+1)}
	=\tfrac{(\sqrt{1+\beta}+\sqrt{1-\beta})(\sqrt{1+\beta c}-\sqrt{1-\beta c})}{(\sqrt{1+\beta}-\sqrt{1-\beta})(\sqrt{1+\beta c}+\sqrt{1-\beta c})}
	\,.
\end{equation}
In terms of new variables the physical region for the annihilation channel is defined by the inequalities
\begin{equation}
	0<\beta<1\,,\quad -1<c<1\,,\quad
	0<\xi<1\,,\quad -1<\chi<1\,.
\end{equation}
Using the new variables and functional relations for the Goncharov's polylogarithms, we have expressed the results of Ref. \cite{LM2019} in terms of the following functions:
\begin{itemize}
	\item $G\left(w_1,.\ldots,w_n|\,\beta \right)$ with $w_i \in\big\{0,\,\pm1,\,\pm\frac{1}{c},\,\pm c+is,\,\pm c-is,\,\pm c+s',\,\pm c-s'\big\}$, where $s=\sqrt{1-c^2},\ s'=\sqrt{3+c^2}$,
	\item $G\left(w_1,.\ldots,w_n|\,2\xi \right)$ with $w_i \in\left\{0,\pm2 i,\pm2,\pm\frac{2 i}{\chi },\pm\frac{2}{\chi }\right\}$,
	\item $G\left(w_1,.\ldots,w_n|\,4\xi^2 \right)$ with $w_i \in\left\{0,\pm4,\pm\frac{4}{\chi },\pm\frac{4}{\chi ^2}\right\}$.
\end{itemize}
Note that the forward-backward symmetry $t\leftrightarrow u$ acts on the new variables as
\begin{equation}
	(\beta,c)\to (\beta,-c)\,,\quad (\xi,\chi)\to (\xi,-\chi)\,.
\end{equation}
The final result contains these polylogarithms up to the fourth weight. Using the results for $\cA_{2\gamma}^{(1)}$ and $\cA_{3\gamma}^{(2)}$ and Eq. \eqref{eq:HviaA} we have obtained the corresponding finite hard amplitudes $\cH_{2\gamma}^{(1)}$ and $\cH_{3\gamma}^{(2)}$.

\section{Results and ancillary files}
The results of the present work are the expressions for invariant amplitudes and related quantities sufficient for construction of the $C$-even part of NNLO differential cross section inclusive with respect to soft photons up to power corrections in electron mass. In particular, the analytical results for the amplitudes $\cA_{3\gamma}^{(2)}$ have been derived here for the first time to the best of our knowledge. Some of results have been rederived, correcting and/or improving if needed the previous results. Below we list the results which we attach to the present paper with explicit statement about their novelty/familiarity:
\begin{enumerate}
	\item Results for the soft-virtual and soft-real functions $V$ and $W$, Eqs. \eqref{eq:SVfunction} and \eqref{eq:Wint}, in dimensional regularization up to $\epsilon^0$ terms. The latter has been obtained earlier in Ref. \cite{Lee:2020zpo} by one of the authors, and here we present the results of that paper for completeness. The result for soft-virtual factor in the most general form is obtained for the first time in the present paper to the best of our knowledge. The high-energy asymptotics of $V$ was presented previously in Ref. \cite{Fadin:2023phc}. Note that, according to the considerations of Ref. \cite{Fadin:2023phc}, higher orders in  $\e$ in both $V$ and $W$ are not needed for the calculation at any loop order. The explicit forms are presented in Eqs. \eqref{eq:Vt}, \eqref{eq:Vs}, \eqref{eq:Wst}. The corresponding ancillary files can be found in \texttt{VW/} folder.

	\item Well-known results for the one- and two-loop polarization operator, see Refs. \cite{KallenSabry,PhysRevD.109.096020}. The corresponding ancillary files can be found in \texttt{PO/} folder.

	\item Form factors and their various asymptotics.
	\begin{itemize}
		\item ``Universal'' contributions from the diagrams $(0)-(2e)$ in Fig. \ref{fig:ff} containing only one lepton kind up to two loops. Those contributions have been obtained earlier in Refs. \cite{Mastrolia2003,Bonciani2004}, however, the printed expressions contain several typos. Nevertheless, we have successfully compared our results with the computer-readable file kindly sent by the authors of \cite{Bonciani2004} and found an exact agreement.\footnote{When comparing, one should keep in mind that the loop measure in that paper differs from the one in this paper.} The results for these contributions together with their various asymptotics are located in \texttt{FF/} folder.
		\item Contribution from the two-loop diagram $(2f)$ in Fig. \ref{fig:ff} with insertion of fermion loop with lepton of another flavor. This contribution has been obtained recently in Ref. \cite{Ahmed2024}, however only in a small subdomain of the whole physical region. We find the agreement of our results with those of Ref. \cite{Ahmed2024}. The results for this contribution together with their various asymptotics are located in \texttt{FFM/} subfolder.
		\item Integration kernels in Eq. \eqref{eq:hadr_vert} for the contributions to the form factors of the diagram with hadronic polarization operator insertion, Fig. \ref{fig:hvpvertex}. Explicit forms are presented in Eqs. \eqref{eq:K1}, \eqref{eq:K2}. Somehow, we were not able to find these explicit formulas in the literature. The results for these integration kernels are located in \texttt{FFH/} folder.
	\end{itemize}
	\item Hard invariant amplitudes $H_{k,n\gamma}^{(L)}$ with $n=1,2,3$ at $\e=0$. They are related to the corresponding amplitudes $A_{k,n\gamma}^{(L)}$ via Eq. \eqref{eq:HviaA}. According to Eq. \eqref{eq:csviaH}, these amplitudes are sufficient for obtaining the observable differential cross section. These amplitudes for $L=2$ are obtained in the present paper for the first time. The results for these amplitudes together with their various asymptotics are located in \texttt{H/} folder.
	Note that, in order to find the high-energy asymptotics of the amplitudes, given the absence of collinear divergences in these contributions, we find it simpler to obtain the fully massless amplitudes by a separate calculation. Besides, it provided an independent cross check of our exact results.
\end{enumerate}
Each folder contains the \texttt{Readme.md} file describing the folder content in detail.

We have performed some additional cross checks of our results. First, we have numerically reproduced the contributions from the diagram $(2f)$ in Fig. \ref{fig:ff} located in \texttt{FFM/} folder using the integration kernels in \texttt{FFH/} and known one-loop polarization operator. Next, we have performed comparison of our results with those of Ref. \cite{Blumlein:2020jrf} where the QED initial state corrections to the process of $e^+e^-$ annihilation into virtual boson have been considered. This comparison, which concerns only the contributions depending on the electron form factor and $\mathcal{W}_{II}$, has shown a perfect agreement.\footnote{One should pay attention to the fact that in Ref. \cite{Blumlein:2020jrf} the restriction on the \emph{total} energy of soft photons was used, while in the present paper the energy of \emph{each} soft photon is restricted.}

\section{Discussion and Conclusion}

Let us now discuss the obtained results. Although the full discussion should include also the polarization effects which can be readily obtained from the invariant amplitudes of the present work, here we will restrict ourselves only to the discussion of the unpolarized differential cross section. Besides, while our results can be equally applied to the $\tau$ pair production, here we will consider only the process \eemumu. The Mathematica notebook \texttt{CrossSection.nb} constructing this quantities from the hard amplitudes and soft factors $V$ and $W$ is attached to the paper. Note that for the numerical results we did not consider the contributions related to hadronic polarization operator.

As it is obvious from our results, the differential cross section is polynomial in two variables, $L_\omega = \ln \frac{\sqrt{s}}{2\omega_0}$ and $L=\ln \frac{s}{m^2}$, with coefficients being the functions of $\beta=\sqrt{1-4M^2/s}$ and $c=\cos\theta$. Namely, we will write
\begin{equation}
	\tfrac{d\sigma_{C\text{-even}}}{d\Omega} =\tfrac{d\sigma_0}{d\Omega}\left[1 +  \delta^{(1)} + \delta^{(2)}\right]\,,
\end{equation}
where
\begin{equation}
	\frac{d\sigma_0}{d\Omega} = \frac{\alpha^2\beta}{4s}\left[2-\beta^2(1-c^2)\right]
\end{equation}
is the differential Born cross section and
\begin{align}
\delta^{(1)} &= \sum_{k,n=0}^{1}\delta^{(1)}_{kn} L_\omega^k L^n,\\
\delta^{(2)} &= \sum_{k,n=0}^{2}\delta^{(2)}_{kn} L_\omega^k L^n
			+\delta^{(2)}_{03} L^3,
			\label{eq:delta2_form}
\end{align}
are the one- and two-loop relative corrections, respectively.
The coefficients  $\delta^{(L)}_{kn}$ are functions of $\beta$ and $c$. Since in the present paper we consider only the $C$-even contributions, $\delta^{(L)}_{kn}$ are \emph{symmetric} with respect to the replacement $c\to -c$.
Note that the coefficients $\delta^{(L)}_{kn}$ with $n>0$ receive contributions from $1\gamma$-reducible diagrams only. The term $\delta^{(2)}_{03} L^3=-a^2\frac89 L^3$ comes from the electron vacuum polarization insertion in the electron form factor.

For the sake of discussion we will split $\delta^{(2)}$ into the sum \begin{equation}
	\delta^{(2)}=\delta^{(2,\text{red})}+\delta^{(2,\text{irr})},
\end{equation}
where $\delta^{(2,\text{red})}$ represents contribution of $1\gamma$-reducible diagrams only as well as the terms which come from the expansion of $\exp[\mathcal{W}_{II}+\mathcal{W}_{FF}]$. Meanwhile $\delta^{(2,\text{irr})}$ represents the remaining contributions involving the $2\gamma$- and $3\gamma$-reducible diagrams or expansion terms of $\exp[\mathcal{W}_{IF}]$.

First, we note that the angular dependence of $\delta^{(2,\text{red})}$ appears entirely due to the muon $F_2$ form factor contribution. However, this form factor is small in comparison to $F_1$ both on the threshold and at high energies. We find that the relative magnitude of the contributions to $\delta^{(2,\text{red})}$  involving $F_2$ form factor is about $10^{-2}$ and the variation of $\delta^{(2,\text{red})}$ is even less than this estimate. So, with sufficient precision we can assume that $\delta^{(2,\text{red})}$ is independent of the angle.

\begin{figure}
	\centering
	\includegraphics[width=0.66\textwidth]{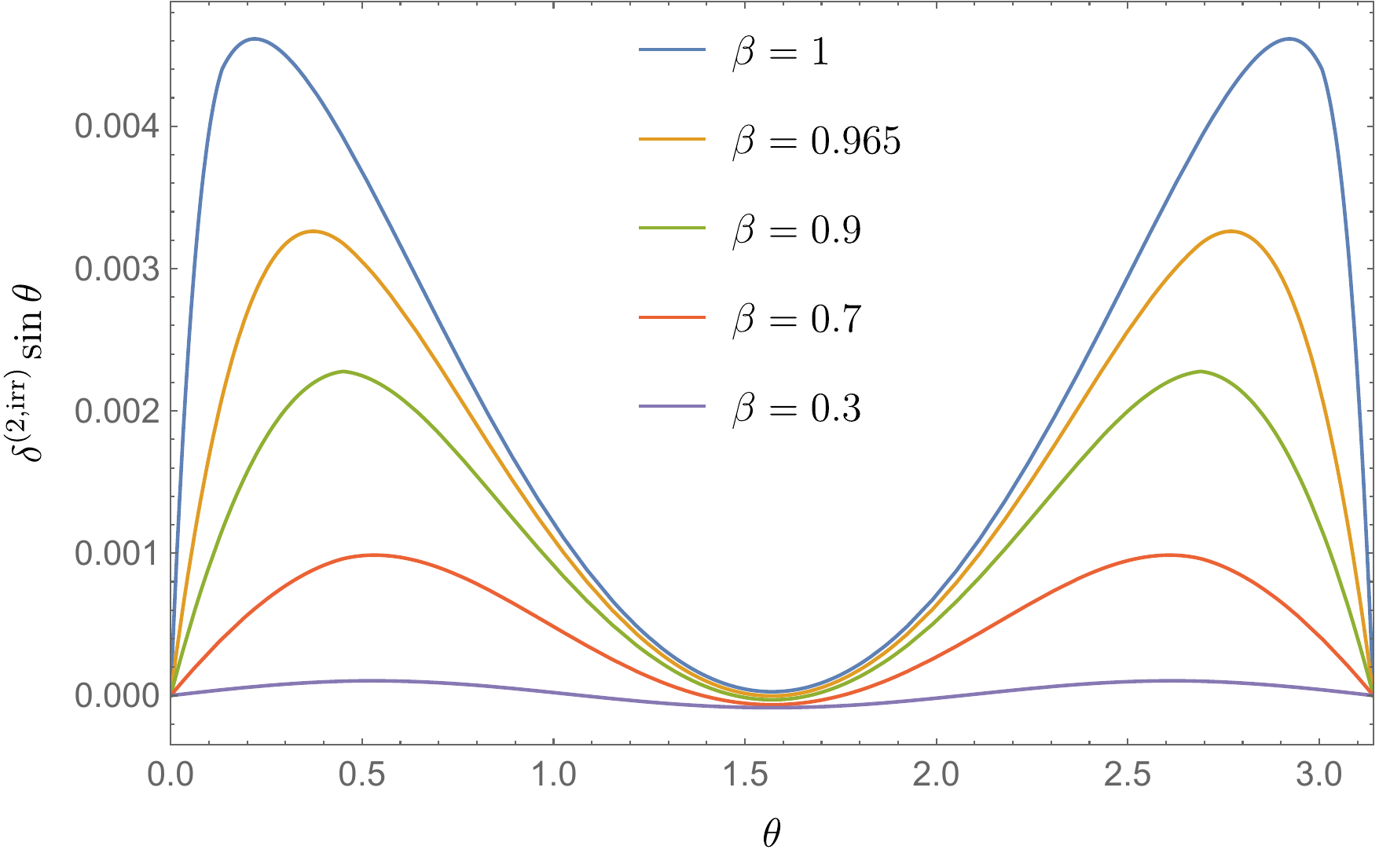}
	\caption{Angular dependence of $1\gamma$-irreducible contribution $\delta^{(2,\text{irr})}$ multiplied by $\sin\theta$. Here we have taken $2\omega_0/\sqrt{s}=0.01$.}
	\label{fig:delta23}
\end{figure}

This is completely different for $\delta^{(2,\text{irr})}$ which angular dependence is shown in Fig. \ref{fig:delta23} for various $\beta$. Note that, for better appearance of the graphs in this figure, we have multiplied  $\delta^{(2,\text{irr})}$ by $\sin\theta$, which should be remembered when judging about the magnitude of this quantity. With the account of this additional $1/\sin \theta$ factor the angular dependence of $\delta^{(2,\text{irr})}$ at small angles proves to be quite sharp. In fact, at high energy and small angles the quantity  $\delta^{(2,\text{irr})}$ can be shown to behave as $\ln^4 (1-\beta^2c^2)$, while its high-energy asymptotics evaluated at $m=M=0$ behaves as $\ln^4 (1-c^2)$ which means that the latter is valid only when $|q_\perp|\gg M$, where $q_\perp$ denotes the component of muon momentum perpendicular to the collision axis.

\begin{figure}
	\centering
	\includegraphics[width=0.66\textwidth]{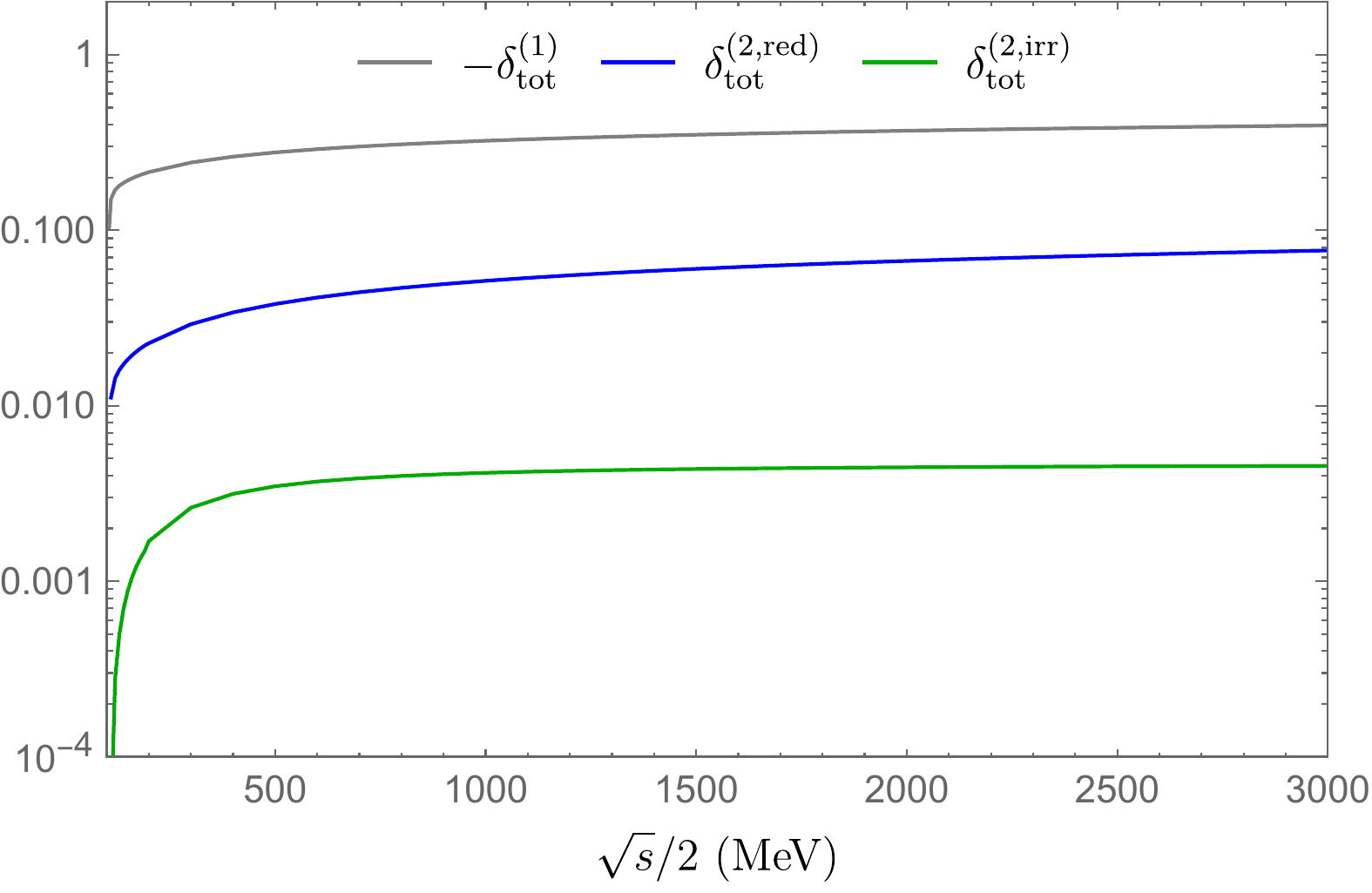}
	\caption{Relative one-loop and two-loop corrections to the total cross section. Here we have taken $2\omega_0/\sqrt{s}=0.01$. Note that the one-loop correction is negative and shown with the minus sign.}
	\label{fig:delta}
\end{figure}

In Fig. \ref{fig:delta} we present corrections to the total cross section,
\begin{equation}
	\sigma_{\text{tot}} =\sigma_{0}\left[1 +  \delta^{(1)}_{\text{tot}} + \delta^{(2,\text{red})}_{\text{tot}}+\delta^{(2,\text{irr})}_{\text{tot}}\right]\,,
\end{equation}
where
\begin{equation}
	\sigma_0=\int d\Omega\frac{d\sigma_0}{d\Omega} = \frac{2\pi\alpha^2\beta}{3s}\left[3-\beta^2\right]
\end{equation}
is the total Born cross section. It can be seen that the relative magnitude of $\delta^{(2,\text{red})}_{\text{tot}}$ can reach a few percent for the chosen $\omega_0$. However, the magnitude $\delta^{(2,\text{irr})}_{\text{tot}}$ in the total cross section always stays below 1\%. One should remember though that the magnitude of $\delta^{(2,\text{irr})}$ in the differential cross section \emph{at small angles} ($\sin{\theta}\ll 1$) may be substantially larger due to its sharp angular dependence, see Fig.\ref{fig:delta23}.

\begin{figure}
	\centering
	\includegraphics[width=0.66\linewidth]{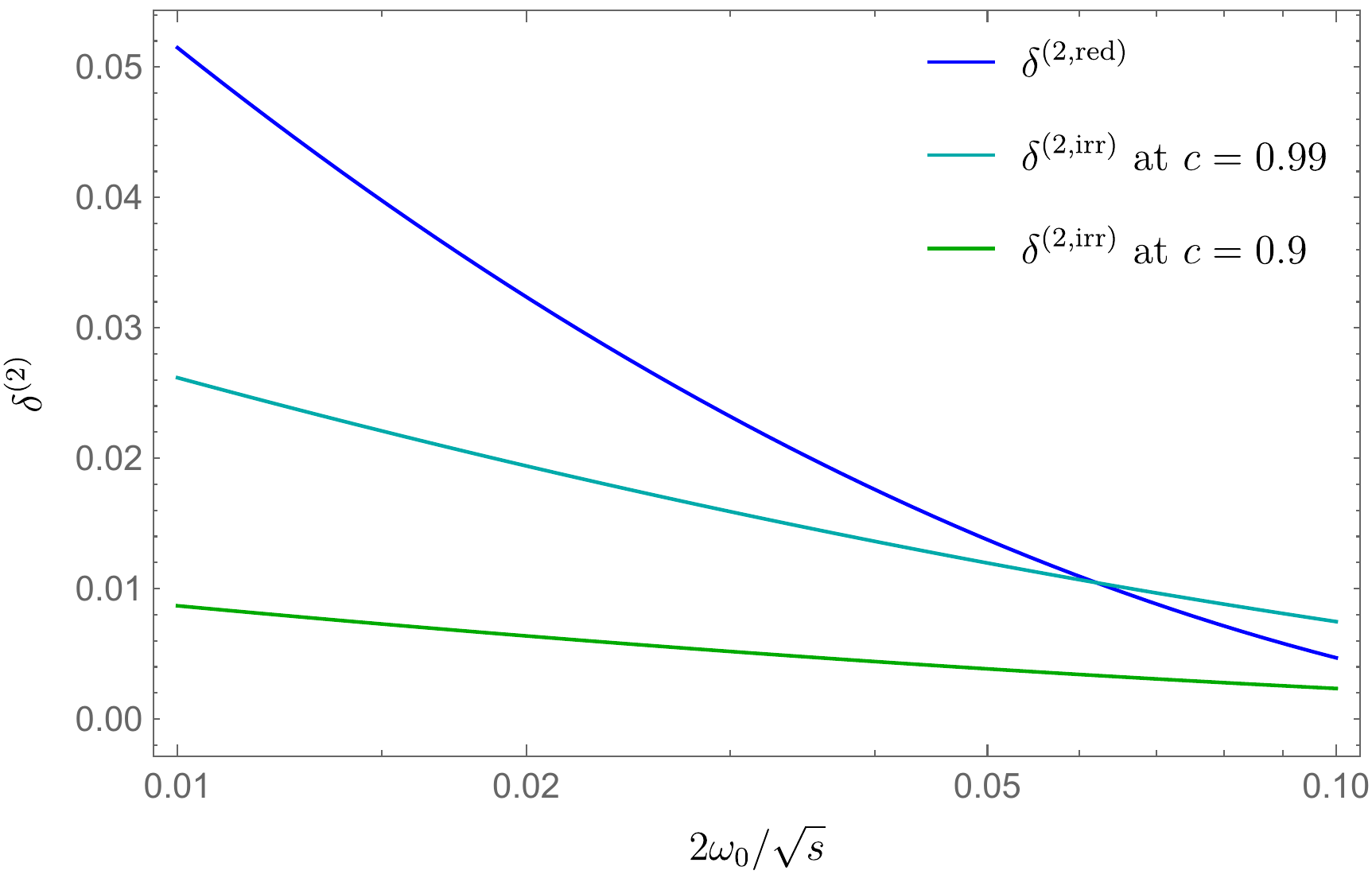}
	\caption{$\delta^{(2,\text{red})}$ and $\delta^{(2,\text{irr})}$ as functions of $2\omega_0/\sqrt{s}$. Here we have taken $\sqrt{s}=2$GeV.}
	\label{fig:omegadep}
\end{figure}

Finally, in Fig. \ref{fig:omegadep} the dependence of $\delta^{(2)}$ on $2\omega_0/\sqrt{s}$ is shown. We see that on the interval $2\omega_0/\sqrt{s}\in [0.01,0.1]$ the dependence is quite strong. In particular, if we take $\omega_0=0.1 \sqrt{s}/2$, the relative magnitude of the two-loop corrections drastically reduces.

To conclude, we have calculated all NNLO contributions to the $C$-even part of the differential cross section for \eemumu process. We used the fact that the ratio $m^2/M^2=m_e^2/m_\mu^2$ is tiny and neglected the power corrections in this parameter. In addition to the two-loop contributions coming from the diagrams with three intermediate photons, which were obtained for the first time, we have rederived also some known results important for our present goal. In particular, we have recalculated the massive lepton QED form factors at two loops and found some typos in the previous papers. Our results can be used for arbitrary polarization of all involved particles.

\acknowledgments
We are grateful to A. Bondar, V. Fadin, and I. Logashenko for stimulating and fruitful discussions. The work has been supported by Russian Science Foundation under grant 20-12-00205.

\appendix

\section{Soft virtual and real functions $V(k_1,k_2)$ and $W(k_1,k_2|\omega_0)$}\label{sec:SV}

Let us calculate the soft virtual function $V(k_1,k_2)$ defined in Eq. \eqref{eq:SVfunction} for general masses of two particles, $k_1^2=m_1^2$, $k_2^2=m_2^2$.
We first calculate this function in the scattering channel, i.e., when one of the incoming momenta has negative time-like component. It corresponds to the condition $-\infty<s<(m_1-m_2)^2$. Besides, we assume that $m_2>m_1$.
Therefore, we calculate
\begin{equation}
	V\left(k_1,-k_2\right)=-8\pi^2\int\frac{\underline{d}^{d}k}{i\left(2\pi\right)^{d}}\frac{1}{k^{2}+i0}\left(\frac{2k_1-k}{k^{2}-2kk_1+i0}
	-\frac{2k_2-k}{k^{2}-2kk_2+i0}\right)^{2}\label{eq:SVfunction1}
\end{equation}
and assume that $k_1^0>0$ and $k_2^0>0$.

It is convenient to define the following variables
\begin{equation}
	x_{12}= \tfrac{k_1\cdot k_2}{m_1m_2} - \sqrt{\left(\tfrac{k_1\cdot k_2}{m_1m_2}\right)^2-1},\quad
	\quad \beta_{12}=
%SV.beta->x
	\frac{1-x_{12}^2}{1+x_{12}^2}
%SV.beta->x/
	=\sqrt{1-\left(\tfrac{m_1m_2}{k_1\cdot k_2}\right)^2}\,.\label{eq:xgb}
\end{equation}
The physical meaning of the variable $\beta_{12}$ is that it is the relative velocity of particles.

We consider the family of integrals
\begin{equation}
	\texttt{SV}(n_1,n_2,n_3) =\int \frac{\underline{d}^dk}{i\pi^{d/2}}[-k^2-i0]^{-n_1}
	[-k^2+2kk_1-i0]^{-n_2}
	[-k^2+2kk_2-i0]^{-n_3}
\end{equation}
and perform the IBP reduction. We find 3 master integrals all expressible in terms of $\Gamma$-functions and hypergeometric $_2F_1$ functions:
\begin{align}
	\frac{\texttt{SV}(0,1,0)}{m_1^{2-2\e}}&=\frac{\texttt{SV}(0,0,1)}{m_2^{2-2\e}}=\Gamma(-1+\e),\\
	\texttt{SV}(0,1,1)&=\frac{\Gamma (-1+\e)x_{12} }{m_1 m_2 \left(1-x_{12}^2\right)}
	\bigg\{m_1^{2-2 \e } \, _2F_1\left(1,2-2 \e ;2-\e ;\tfrac{x_{12}\, \left(m_1-m_2 x_{12}\right)}{m_2 \left(1-x_{12}^2\right)}\right)
	\nonumber\\
	&-m_2^{2-2 \e } \, _2F_1\left(1,2-2 \e ;2-\e ;\tfrac{m_1-m_2 x_{12}}{m_1(1-x_{12}^2)}\right)\bigg\}\,.
\end{align}

The function $V(k_1,-k_2)$ is expressed via these master integrals as
\begin{multline}
	V(k_1,-k_2)=
	\left(\tfrac{m_2/m_1-1}{2(k_1\cdot k_2-m_1m_2)}
	-\tfrac{m_2/m_1+1}{2(k_1\cdot k_2+m_1m_2)}
	+\tfrac{2-\epsilon}{2 m_1^2}\right) (1-1/\e)
	\texttt{SV}(0,1,0)
	\\
	+\left(\tfrac{m_1/m_2-1}{2(k_1\cdot k_2-m_1m_2)}
	-\tfrac{m_1/m_2+1}{2(k_1\cdot k_2+m_1m_2)}
	+\tfrac{2-\epsilon}{2 m_2^2}\right) (1- 1/\e)
	\texttt{SV}(0,0,1)
	\\
	-\bigg(\tfrac{(2-1/\e)(k_1-k_2)^2}{  2(k_1\cdot k_2-m_1m_2)}
	+\tfrac{(2-1/\e)(k_1-k_2)^2}{2(k_1\cdot k_2+m_1m_2)}+1\bigg)\texttt{SV}(0,1,1)\,.
\end{multline}

Substituting and expanding in $\e$, we obtain
\begin{multline}
	V(k_1,-k_2)=-2\left(m_1m_2\right)^{-\e}\bigg\{
%SV.Vt
	\frac{1+\frac1{\beta_{12}}\ln{x_{12}}}{\e}+1
	-\frac{\left(1-x_{12}^2\right)m_1 m_2}{2t_{12} x_{12}} \ln{x_{12}}+\frac{m_2^2-m_1^2}{2t_{12}}\ln {\tfrac{m_1}{m_2}}
	\\
	+\frac{1}{\beta_{12}}\left[ f\left(x_{12}^2\right)-f\left(\tfrac{m_1 x_{12}}{m_2}\right)-f\left(\tfrac{m_2 x_{12}}{m_1}\right)\right]
%SV.Vt/
	+O(\e)\bigg\}
	\label{eq:Vt}
	\end{multline}
where $t_{12}=(k_1-k_2)^2=
%SV.t->x
(m_1x_{12}-m_2) (m_1 - m_2 x_{12})/x_{12}
%SV.t->x/
$ and
\begin{equation}
	f\left(z\right)=
	\tfrac12\left[\mathrm{Li}_{2}\left(1-z\right)-\mathrm{Li}_{2}\left(1-z^{-1}\right)\right]=
%SV.f(z)
	\mathrm{Li}_{2}\left(1-z\right)+\tfrac{1}{4}\ln^2{z}
%SV.f(z)/
\,.
\end{equation}
Note that this formula is valid for any ratio $m_1/m_2$ provided that $x_{12}\in (0,1)$. In order to perform the analytical continuation to the annihilation channel, we should follow a path $C$ in the upper half-plane of the complex variable $t$.
We see that on this path $t\stackrel{C}{\rightarrow} s=(k_1+k_2)^2$, $\beta_{12}\stackrel{C}{\rightarrow} \beta_{12}$ and $x_{12}\stackrel{C}{\rightarrow} -x_{12}+i0$. However, a special care is required for the analytical continuation of the function $f(x_{12}^2)$ as its argument winds around the origin. Expressing it in terms of the function $\mathrm{Li}_2(x_{12})$ and logarithms, one can check that the following substitution rule holds
\begin{equation}
	f(x_{12}^2) \stackrel{C}{\longrightarrow} f(x_{12}^2)-\pi^2-2i\pi\ln{\frac{1-x_{12}^2}{x_{12}}}
\end{equation}

\begin{multline}
	V(k_1,k_2)=-2\left(m_1m_2\right)^{-\e}\bigg\{
%SV.Vs
	\frac{1+\frac1{\beta_{12}}\ln(-x_{12}+i0)}{\e}
	+1+\tfrac{\left(1-x_{12}^2\right)m_1 m_2}{2s_{12}\,x_{12}} \ln(-x_{12}+i0)
	\\
	+\tfrac{m_2^2-m_1^2}{2s_{12}}\ln {\tfrac{m_1}{m_2}}+\frac{1}{\beta_{12}}\left[ f\left(x_{12}^2\right)-\pi^2-2i\pi\ln{\tfrac{1-x_{12}^2}{x_{12}}} -f\left(-\tfrac{m_1 x_{12}}{m_2}+i0\right)-f\left(-\tfrac{m_2 x_{12}}{m_1}+i0\right)\right]
%SV.Vs
	\\+O(\e)\bigg\}
	\label{eq:Vs}
\end{multline}
where $x_{12}$ and $\beta_{12}$ are defined in \eqref{eq:xgb} and $s_{12}=(k_1+k_2)^2=
%SV.s->x
(m_1x_{12}+m_2) (m_1 + m_2 x_{12})/x_{12}
%SV.s->x/
$.

For our present goal we need the following special cases of the obtained formulae. First, we need $V(p_1,-q_1)$ and $V(p_1,-q_2)$. Assuming that $m^2\ll M^2\ll |t|$, we obtain
\begin{multline}
	V(p_1,-q_1) = -2 M^{-2 \e } \bigg\{
%SV.Vt.asy
	\frac{-\ln {\frac{M^2-t}{M^2}}-\ln{\tfrac{M}{m}}+1}{\e }
	\\
	-\text{Li}_2\left(\tfrac{-t}{M^2-t}\right)
	+\tfrac{1}{2} \ln^2 {\tfrac{M^2-t}{M^2}}
	+\tfrac{M^2-t}{2t}\ln {\tfrac{M^2-t}{M^2}}
	-\ln ^2{\tfrac{M}{m}}+\tfrac{1}{2} \ln{\tfrac{M}{m}}+1
%SV.Vt.asy/
	\bigg\}
\end{multline}
and $V(p_1,-q_2)$ has the same form with $t\to u$.

Note that in the difference $V(p_1,-q_1)-V(p_1,-q_2)$ the terms independent of $t$ and $u$ cancel out, so that we have
\begin{multline}
	\mathcal{V}_{IF} = 2[V(p_1,-q_1)-V(p_1,-q_2)]= -4 M^{-2 \e } \bigg\{\frac{\ln {\frac{M^2-u}{M^2-t}}}{\e }
	-\text{Li}_2\left(\tfrac{-t}{M^2-t}\right)+\text{Li}_2\left(\tfrac{-u}{M^2-u}\right)
	\\
	+\tfrac{1}{2} \ln^2 {\tfrac{M^2-t}{M^2}}-\tfrac{1}{2} \ln^2 {\tfrac{M^2-u}{M^2}}
	+\tfrac{M^2-t}{2t}\ln {\tfrac{M^2-t}{M^2}}-\tfrac{M^2-u}{2u}\ln {\tfrac{M^2-u}{M^2}}
	\bigg\}
\end{multline}

Next, we need $V(-q_1,-q_2)=V(q_1,q_2)$ for $m_1=m_2=M$. We have
\begin{multline}
	\mathcal{V}_{FF} = V(q_1,q_2)=	-2 M^{-2\e}\bigg\{
%SV.Vseq
	\frac{1+\frac{1+x_s^2}{1-x_s^2}(\ln{x_s}+i \pi
			)}{\e}
+1+\frac{1-x_s }{2 (1+x_{s})}(\ln{x_{s}}+i \pi )
\\
+\frac{1+x_s^2}{1-x_s^2}\left[2
	\text{Li}_2(1-x_{s})+\tfrac{1}{2}\ln ^2{x_{s}}-i \pi  \ln{\frac{(1-x_{s})^2}{x_{s}}}- \pi ^2\right]
%SV.Vseq/
	+O(\e)\bigg\}\,,
\end{multline}
where $x_s=\frac{1-\beta}{1+\beta}$.

Finally, let us present for completeness also the expression for soft real function $W$ obtained in Ref. \cite{Lee:2020zpo}.
Defined as in Eq. \eqref{eq:Wint},
\begin{equation}
	W(k_1,k_2|\omega_0)=-16\pi^2\intop_{\omega<\omega_0}\frac{\underline{d}^{d-1}k}{(2\pi)^{d-1}2\omega}\left(\frac{k_1}{k\cdot k_1}-\frac{k_2}{k\cdot k_2}\right)^2, \label{eq:Wint1}
\end{equation}
this function evaluates to
\begin{multline}
	W(k_1,k_2|\omega_0)=-4\left(2\omega_0\right)^{-2\e}
	\bigg\{
%SV.W
	-\frac{1+\frac1{\beta_{12}}\ln {x_{12}}}{\e}+
	\frac1{\beta_{12}}\bigg[f\left({x_{12}\, x_1}/{x_2}\right)+f\left({x_{12}\, x_2}/{x_1}\right)\\
	-f\left({x_1 x_2}/{x_{12}}\right)+f\left(x_{12}\, x_1 x_2 \right)
	-f\left(x_{12}^2\right)\bigg]
	+\frac{1}{\beta_1} \ln {x_1}+\frac{1}{\beta_2} \ln {x_2}
%SV.W/
	+O(\e)\bigg\}\,,
	\label{eq:Wst}
\end{multline}
where $\beta_{12}=\sqrt{1-\left(m_1m_2/k_1\cdot k_2\right)^2}$ is the relative velocity and $\beta_i=\sqrt{1-\left(m_i/k_i^0\right)^2}$ ($i=1,2$) are the velocities of particles in the lab frame, $x_k=\sqrt{\tfrac{1-\beta_k}{1+\beta_k}}$.

\section{Contribution to the form factors from the insertion of vacuum polarization of another lepton flavor}\label{sec:vp_vertex}

We consider the contribution to the two-loop lepton form factors of the diagram $(2f)$ in Fig.~\ref{fig:ff}. In Ref. \cite{Ahmed2024} this contribution has been already considered in the scattering channel in the region
\begin{equation}
	m_i > m,\quad q^2 <4(m^2 - m_i^2).
\end{equation}
We have independently obtained this contribution for the whole physical region including the annihilation channel which is needed for our main goal in the present paper.
We define the integral family as follows:
\begin{equation}
	\tilde{I}_{n_1,\ldots,n_7} = \int \frac{dl_1 dl_2}{(i\pi^{d/2})^2} \prod_{i=1}^{7} (\tilde{D}_i - i0)^{-n_i},
\end{equation}
where
\begin{gather*}
	\tilde{D}_1 = m^2 - (p_1 - l_1)^2,\quad
	\tilde{D}_2 = m^2 - (p_2 - l_1)^2,\quad
	\tilde{D}_3 = m_i^2 - l_2^2,\\
	\tilde{D}_4 = m_i^2 - (l_1 + l_2)^2,\quad
	\tilde{D}_5 = -l_1^2,\quad
	\tilde{D}_6 = l_2 p_2,\quad
	\tilde{D}_7 = l_2 p_1.
\end{gather*}

Using \LiteRed \cite{Lee2013a,LeeLiteRed2} for IBP reduction, we identify 7 master integrals:
\begin{gather*}
	(\tilde{j}_1, \ldots, \tilde{j}_7) = (\tilde{I}_{0011000}, \tilde{I}_{0101000}, \tilde{I}_{0111000}, \tilde{I}_{0211000}, \tilde{I}_{1101000}, \tilde{I}_{1111000}, \tilde{I}_{2111000}).
\end{gather*}

This integral family depends on two dimensionless ratios, $q^2/m^2$ and $m_i/m$. To reduce the differential system to $\epsilon$-form, we introduce the variables $x$ and $z$, defined as:
\begin{equation}
	\frac{q^2}{m^2} = -\frac{(1-x)^2}{x},\quad \frac{m_i}{m} = \frac{ (1+x)z}{z^2 + x}.
\end{equation}
\begin{figure}
	\centering
	\includegraphics[width=0.5\linewidth]{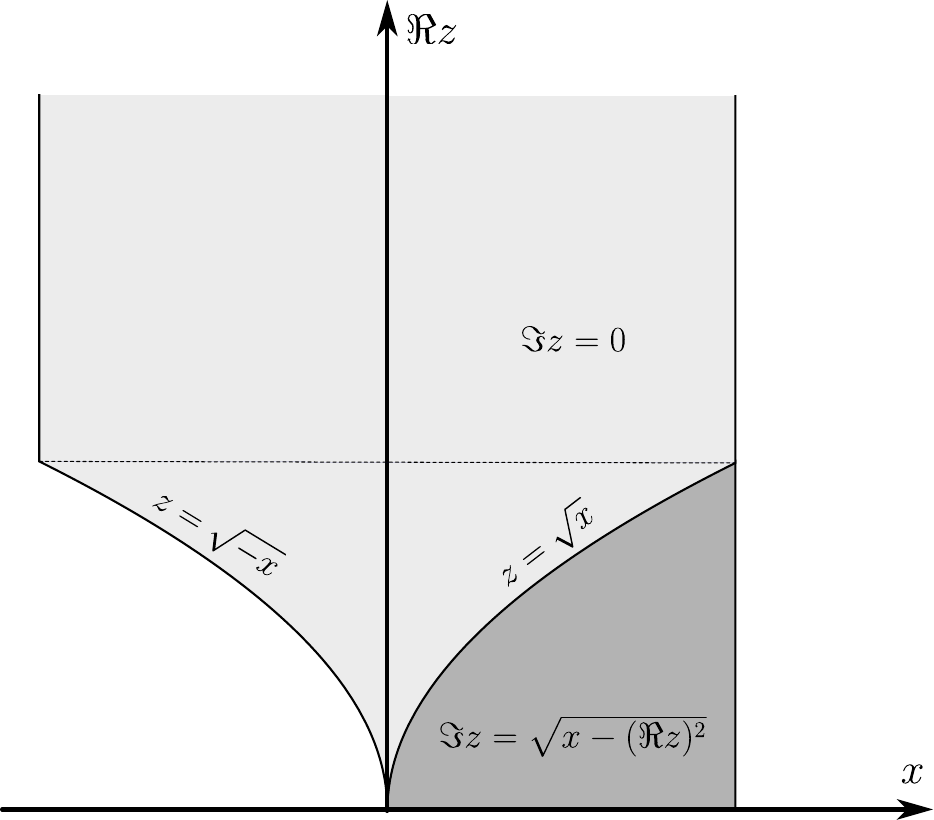}
	\caption{Physical region in terms of variables $x$ and $z$.}
	\label{fig:xzregion}
\end{figure}

The physical region in terms of the variables $x$ and $z$ is shown in Fig. \ref{fig:xzregion}. The part of the physical region where $z$ becomes complex is shown in darker color. Note that the region considered in Ref. \cite{Ahmed2024} corresponds to a curvilinear triangle restricted by the inequalities
\begin{equation}
	0<x<z^2<1\,.\label{eq:regMA}
\end{equation}
We first restrict ourselves to the region $0 < x < 1$ and $z > 1$. The results for whole physical region are obtained via analytical continuation.

Using \Libra \cite{Lee:2020zfb}, we reduce the system to $\e$-form \cite{Henn2013, Lee:2014ioa}. We fix the boundary conditions by considering the asymptotics $x\to 1$ and $z\to \infty$. More specifically, we first consider the limit $\bar{x}=1-x \to 0$ and determine which coefficients of the asymptotic expansion in $\bar{x}$ are to be calculated. Those coefficients still depend on $z$ and we construct the differential system for
them. We reduce this system to $\epsilon$-form and fix the boundary conditions from $z\to\infty$ asymptotics using the expansion by regions. As a result, we obtain the master integrals in terms of Goncharov's polylogarithms with arguments $\bar{x}$ and $z^{-1}$.

The renormalization of these form factors requires the account  of two diagrams with counterterms. The first is the one-loop diagram with a one-loop polarization operator counterterm. The second is the tree diagram with a two-loop vertex counterterm depending on two different masses. This contribution can be accounted for by subtracting the obtained form factor at $q^2=0$. Alternatively, we can use the known result from \cite{Davydychev1999} for the corresponding counterterm. The renormalized form factors contain no infrared divergences, as expected.

To simplify the result and, especially, to perform the analytical continuation, it is convenient to rewrite it in terms of ordinary polylogarithms up to third order using the technique described in \cite{Lee2024}. Note that in the physical region the differential system for the master integrals, in addition to the singularity at $x=z^2$ also contains singularity at $z=1$, as shown by the dashed line, however, the specific solution does not contain this spurious branching locus, which serves as a good check of our setup.
We perform the analytical continuation to the region $q^2 > 0$ by following a path in the upper half-plane of $q^2$. It is straightforward to check that $\Im(q^2(x)) > 0$ when $x$ lies within a half-disk. Therefore, the analytical continuation to the region $q^2 > 0$ can be achieved along the path $x = |x| e^{i\phi}$, with $|x|$ fixed and $\phi$ varying from $0$ to $\pi$.

Following a similar strategy, we have succeeded to obtain a universal form of the result valid for the whole physical region. When comparing our results with those of Ref. \cite{Ahmed2024} in the region \eqref{eq:regMA}, we find that our results agree up to an opposite overall sign.

\section{Contribution to the form factors from the hadronic polarization insertion}\label{sec:hadr_vert}

Let us present  the integral kernels $K_{1,2}(y,y_1)$ in Eq. \eqref{eq:hadr_vert}, which reads
\begin{equation}
	F_{k,(\ell)}^{(\text{had})}(s)=-\frac{4a}{\pi}  \int_{s_0}^{\infty } \, \frac{ds_1 }{s_1} K_{k}\left(\tfrac{s}{m_\ell^2},\tfrac{s_1}{m_\ell^2}\right)\Im\Pi^{(\text{had})}(s_1)
\end{equation}

For brevity we will put $m_\ell=1$ below. Then we have

\begin{multline}
	K_1(s,s_1)=
%hvp_vert_K1
	\left[\frac{\left(8+s\right)s_1^2}{2\left(4-s\right)^2s}-\frac{s_1}{4-s}+\frac{s-2}{2s}\right]\frac{1-x}{1+x}g\left(x,x_1\right)
	+\left[\frac{8-3s}{4s}+\frac{\left(8+s\right)s_1}{2\left(4-s\right)s}\right]\frac{1-x}{1+x}\ln{x}\\
	+\left[\frac{\left(16-22s+3s^2\right)s_1^2}{8\left(4-s\right)^2}+\frac{\left(32-5s\right)s\,s_1}{4\left(4-s\right)^2}-\frac{4+s}{2\left(4-s\right)}\right]\frac{1-x_1}{1+x_1}\ln (-x_1)\\
	+\left[\frac{\left(16-22s+3s^2\right)s_1^2}{8\left(4-s\right)^2}+\frac{s\,s_1}{2\left(4-s\right)}\frac{1}{2}\right]\ln{s_1}
	-\frac{\left(8-3s\right)s_1}{4\left(4-s\right)}-1
%hvp_vert_K1/
	\,,\label{eq:K1}
\end{multline}
\begin{multline}
	K_2(s,s_1) =
%hvp_vert_K2
	\left[\frac{4s_1}{\left(4-s\right)s}-\frac{6s_1^2}{\left(4-s\right)^2s}\right]\frac{1-x}{1+x}g\left(x,x_1\right)
	+\left[\frac{1}{s}-\frac{6s_1}{\left(4-s\right)s}\right]\frac{1-x}{1+x}\ln{x}\\
	+\left[\frac{\left(10-s\right)s_1^2}{2\left(4-s\right)^2}-\frac{s_1}{4-s}\right]\frac{1+x_1}{1-x_1}\ln(-x_1)
	-\left[\frac{2s_1}{4-s}-\frac{\left(10-s\right)s_1^2}{2\left(4-s\right)^2}\right]\ln{s_1}
	-\frac{s_1}{4-s}
%hvp_vert_K2/
	\,,\label{eq:K2}
\end{multline}
Here
\begin{equation}
	x=
%hvp_vert_x
	1-\frac{s}{2}\left(1 - \sqrt{1-{4}/{s}}\right)
%hvp_vert_x/
	+i0,\qquad
	x_1=
%hvp_vert_x1
	1-\frac{s_1}{2}\left(1 - \sqrt{1-{4}/{s_1}}\right)
%hvp_vert_x1/
\,,
\end{equation}
and
\begin{equation}
	g(x,x_1)=
%hvp_vert_g
	\text{Li}_2\left(1+\tfrac{x_1}{x}\right)-\text{Li}_2\left(1+xx_1\right)-2\ln\left(1-x_1\right)\ln{x}
%hvp_vert_g/
\end{equation}
It is worth to note that $g(x,x_1) = -g(x^{-1},x_1)=g(x,x_1^{-1})$.

We underline that the above formulae correctly describe the kernels both in the scattering region $s<0,\ s_1>0$ and in the annihilation region $s>4,\ s_1>0$. In particular, when $s_1\in [0,4]$ the square root $\sqrt{1-{4}/{s_1}}$ in the definition of $x_1$ can be understood as $\pm i \sqrt{{4}/{s_1}-1}$ with any sign.

\bibliographystyle{JHEP}
\bibliography{eemumu}
\end{document}